\documentclass[aps,prb,twocolumn,preprintnumbers,amsmath,amssymb,superscriptaddress]{revtex4}

\usepackage{graphicx}
\usepackage{dcolumn}
\usepackage{bm}
\usepackage{color}

\begin{document}

\title{Measurement of nonlinear frequency shift coefficient 
in spin-torque oscillators based on MgO tunnel junctions}

\author{Kiwamu Kudo}
\email{kiwamu.kudo@toshiba.co.jp}
\author{Tazumi Nagasawa}
\author{Rie Sato}
\author{Koichi Mizushima}
\affiliation{Corporate Research and Development Center, 
Toshiba Corporation, Kawasaki, 212-8582, Japan}

\date{\today}

\begin{abstract}
The nonlinear frequency shift coefficient, 
which represents the strength of the transformation of 
amplitude fluctuations into phase fluctuations of an oscillator, 
is measured for MgO-based spin-torque oscillators 
by analyzing the current dependence of the power spectrum. 
We have observed that linewidth against inverse normalized power plots 
show linear behavior below and above the oscillation threshold 
as predicted by the analytical theories for spin-torque oscillators. 
The magnitude of the coefficient is determined from the ratio of the linear slopes. 
Small magnitude of the coefficient ($\sim 3$) has been obtained for 
the device exhibiting narrow linewidth ($\sim 10~\mathrm{MHz}$) at high bias current. 
\end{abstract}


\maketitle

Spin-torque oscillators (STOs) emit a microwave signal. 
The signal originates in magnetization oscillations excited by 
bias dc current in a magnetoresistive (MR) device\cite{Kiselev,Rippard}. 
In recent years, extensive studies have been carried out on STOs 
because they are a promising candidate 
for an on-chip microwave oscillator\cite{Katine,Silva}. 
One of the important properties of an STO is its frequency nonlinearity, 
i.e., a frequency depends on an oscillation amplitude\cite{Slavin,Rezende}. 
Due to the property, 
the frequency of STO is tunable only by changing bias dc current, 
which is generally considered to be an advantage for applications. 
The nonlinearity is, however, a disadvantage of STO 
when thermal fluctuations are taken into account. 
According to the analytical theory of Kim {\it et al.}\cite{Kim}, 
amplitude fluctuations are transformed into phase fluctuations 
because of the nonlinearity, 
resulting in spectrum linewidth broadening.
The linewidth is a measure of the phase stability of oscillation 
and it is preferable that it be narrow. 
Estimating quantitatively the nonlinearity, which determines the device performances, 
is therefore a key subject for further developments of STOs. 
According to the recent theories\cite{Kim,Tiberkevich,Kim2,Tiberkevich3,Kudo,Slavin2}, 
the quantity representing the degree of the nonlinearity 
is the normalized dimensionless nonlinear frequency shift coefficient $\nu$ 
(regarding the definition of the coefficient, 
see, e.g., Eq.~(7) of Ref.~\onlinecite{Tiberkevich3} 
or Eq.~(9) of Ref.~\onlinecite{Kudo}). 
In this letter, we report experimental estimations of the coefficient $\nu$ 
in MgO-based STOs near threshold, 
which have not been directly addressed by previous experiments. 
Following some theoretical remarks, 
experimental results are shown.

\begin{figure}[t]
  \includegraphics{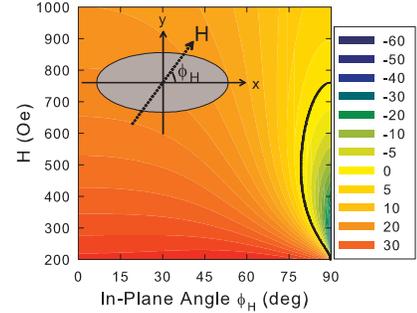}
  \caption{\label{Figure:1}
  (Color online) Dependence of $\nu$ 
  in the planar device with uniaxial anisotropy ($x$-direction) on 
  the magnitude $H$ and the in-plane angle $\phi_H$ of in-plane external magnetic field. 
  The parameters chosen are an uniaxial anisotropy field $H_k=200~\mathrm{Oe}$, 
  a demagnetizing field $4 \pi M_\mathrm{s}=8000~\mathrm{Oe}$, 
  the Gilbert damping $\alpha_\mathrm{G}=0.01$, and 
  the nonlinearity of damping\cite{Tiberkevich2} $q_1=3$. 
  In the notation used here, positive and negative $\nu$ denotes the {\it red} and 
  {\it blue} frequency shift, respectively. 
  }
\end{figure}

It is theoretically known that 
the coefficient $\nu$ has various values depending 
on magnetic environments and damping\cite{Tiberkevich3,Kudo,Slavin2}. 
In general, the coefficient $\nu$ also depends on bias current
\cite{Tiberkevich3,Slavin2}. 
Considering that the variation of $\nu$ with bias current is small 
in the range near threshold, 
we treat $\nu$ as a constant independent of bias current.

First, we demonstrate numerically how large the coefficient $\nu$ is in typical STOs. 
A calculation example of $\nu$ at the threshold 
in a planar device with uniaxial anisotropy is shown in Fig.~\ref{Figure:1}, 
representing the dependence on the magnitude and the angle of an in-plane external magnetic field. 
The calculation is performed by the method described in Ref.~\onlinecite{Kudo} 
which is based on the macrospin model. 
We have used the typical STO parameters shown in the figure caption. 
In the wide external field region, $|\nu|$ is much larger than unity 
($|\nu|\sim 5$-$50$). 
On the black line shown in Fig.~\ref{Figure:1}, the frequency nonlinearity vanishes 
(the nonlinearity due to the demagnetizing effect cancels out that due to 
the in-plane anisotropy), where remarkable reduction of linewidth 
is expected\cite{Slavin2,Thadani,Mizushima}.

\begin{figure}[t]
  \includegraphics{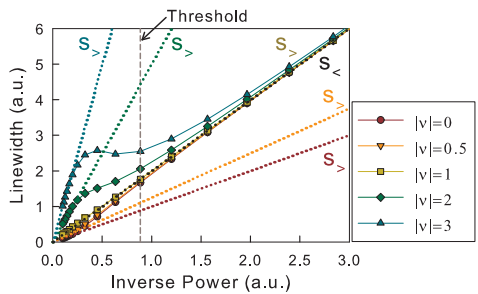}
  \caption{\label{Figure:2}
  (Color online) Linewidth and inverse power near threshold 
  for several values of $|\nu|$. 
  The units of linewidth and power are appropriately scaled. 
  }
\end{figure}

To estimate the value of $\nu$ from experiment, 
we have used the spectrum analysis method based on the theories 
of spectrum linewidth of STO under thermal fluctuations
\cite{Kim,Tiberkevich,Kim2,Tiberkevich3,Kudo,Slavin2}. 
The method is similar to the one often used 
to measure the linewidth enhancement factor ($\alpha$-factor) in lasers\cite{Toffano97}. 
According to the theories, the linewidth shows asymptotic behavior 
below and above threshold. 
In the region below threshold (the thermal activated oscillation region), 
the linewidth 
is given by 
\begin{equation}\label{Eq:LWbelow}
\Delta f^{<}=\Delta f_0 \times (k_\mathrm{B} T\slash E_\mathrm{osci}). 
\end{equation}
In the region above threshold (the current-induced oscillation region), 
the linewidth is given by 
\begin{equation}\label{Eq:LWabove}
\Delta f^{>}=\Delta f_0 \times (k_\mathrm{B} T\slash E_\mathrm{osci})
\times (1+\nu^2)\slash 2, 
\end{equation}
in which the additional phase diffusion 
due to amplitude fluctuations is expressed by the factor of $\nu^2$. 
In Eqs.~(\ref{Eq:LWbelow}) and (\ref{Eq:LWabove}), 
$\Delta f_0$ is the damping at thermal equilibrium expressed in linewidth, 
$k_\mathrm{B} T$ is the thermal energy, and 
$E_\mathrm{osci}$ is the average magnetization oscillating energy 
of a certain oscillation mode. 
The energy $E_\mathrm{osci}$ becomes large as the bias current $I$ increases. 
Since the voltage oscillation signal from STO originates in the MR effect, 
$E_\mathrm{osci}$ is proportional to a `normalized' power $P\slash I^2$, 
where $P$ is the power of voltage oscillation signal from STO. 
Therefore, Eqs.~(\ref{Eq:LWbelow}) and (\ref{Eq:LWabove}) both indicate that 
the linewidth is proportional to the inverse normalized power 
in the two asymptotic regions, 
i.e., $\Delta f^{<,>} \propto (P\slash I^2)^{-1}$. 
Accordingly, drawing a linewidth versus inverse normalized power plot, 
we can extract $\nu$ from the following relation; 
\begin{equation}\label{eq:relation}
s_{>}\slash s_{<}=(1+\nu^2)\slash 2, 
\end{equation}
where $s_{<}$ and $s_{>}$ are the slopes of asymptotes 
below and above threshold, respectively. 
To visualize Eq.~(\ref{eq:relation}), we show 
the linewidth and inverse power calculated numerically in Fig.~\ref{Figure:2}. 
The calculation is based on the Fokker-Planck equation (FPE) 
corresponding to a noisy nonlinear auto-oscillator\cite{Kim2,Seybold}. 
We have used the single-Lorentzian approximation for a power spectrum. 
In the approximation, the linewidth is given by 
the real part of the first eigenvalue in the eigenmode expansion method for FPE
($\mathrm{Re} \Lambda_{1,1}$ in the notation of Ref.~\onlinecite{Kim2}). 
In Fig.~\ref{Figure:2}, linewidths show asymptotic behavior (dotted lines)
represented by Eqs.~(\ref{Eq:LWbelow}) and (\ref{Eq:LWabove}) 
in the region below and above threshold, and 
the relation between the slopes of asymptotes (Eq.~(\ref{eq:relation})) holds.

Measurements are performed 
on MgO-based planar magnetic tunnel junction devices 
at room temperature. 
Tunnel junctions are composed of 
IrMn(10)\slash PL\slash MgO(1.05)\slash FL, 
in which the free layer (FL) is a CoFeB(3) layer and the pinned layer (PL) is a 
CoFe(4)\slash Ru(0.95)\slash CoFeB(4) synthetic antiferromagnet trilayer. 
The junctions have been fabricated by sputter deposition 
with annealing for one hour in a high magnetic field 
and at the temperature of $300^\circ \mathrm{C}$. 
Nanopillar devices are patterned 
using electron-beam lithography and ion milling. 
On the same wafer, we find samples with various TMR ratio 
in the range $10$-$90\%$. 
We have observed that lower TMR ratio samples tend 
to show narrower spectral linewidth oscillations. 
This tendency in MgO-based samples is similar to that reported in Ref.~\onlinecite{Houssameddine08}. 
We consider that the difference of TMR ratio among samples 
may result from localized defects in the MgO barrier 
as speculated by the authors of Ref.~\onlinecite{Houssameddine08}. 
In this letter, we show the results of one typical low TMR ratio sample with 
a $200 \times 120~\mathrm{nm}^2$ elliptical shape. 
From $R$-$H$ (resistance-field) characteristic measurements 
of the sample for several field angles, 
we find that the resistance is 
$235~\mathrm{\Omega}$ ($5.6~\mathrm{\Omega}\cdot \mathrm{\mu m}^2$) 
for parallel 
configuration, TMR ratio is $15.6~\%$, 
and an in-plane anisotropy field $H_k$ is about $200~\mathrm{Oe}$. 
The pinned layer magnetization is along the long axis of the ellipse 
without an applied field.

Oscillation properties are measured by a spectrum analyzer. 
We apply dc current $I$ to the sample 
with antiparallel magnetization configuration. 
Positive $I$ corresponds to electrons flowing from the pinned layer to the free layer. 
Oscillation properties vary sensitively with 
the field applied in the sample plane: 
the magnitude $H$ of $200$-$700~\mathrm{Oe}$ 
and the angle $\phi_H$ of $0$-$90^\circ$. 
When the angle $\phi_H \simeq 40^\circ \pm 5^\circ$ and 
the magnitude $H \simeq 500 \pm 50~\mathrm{Oe}$, 
distinct oscillation peaks are observed. 
In the other applied fields, 
distinct peaks are not observed at any bias current ($I \le 0.7~\mathrm{mA}$). 
Below, we show the results for the two setups of the applied fields 
in which narrow linewidths 
of $\sim 10~\mathrm{MHz}$ are observed: 
(I) $H=490~\mathrm{Oe}$ and $\phi_H=42.4^\circ$, 
and (II) $H=516~\mathrm{Oe}$ and $\phi_H=39.7^\circ$.

\begin{figure}[t]
  \includegraphics{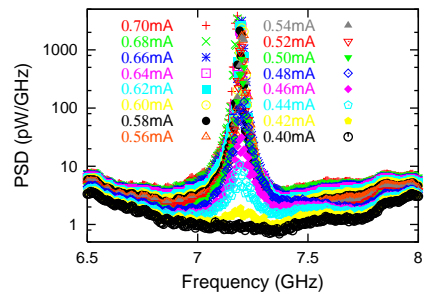}
  \caption{\label{Figure:3}
  (Color online) 
  PSD of the voltage oscillation signal in the setup (I) for 
  bias currents from $0.4$ to $0.7~\mathrm{mA}$. 
  }
\end{figure}

\begin{figure}[t]
  \includegraphics{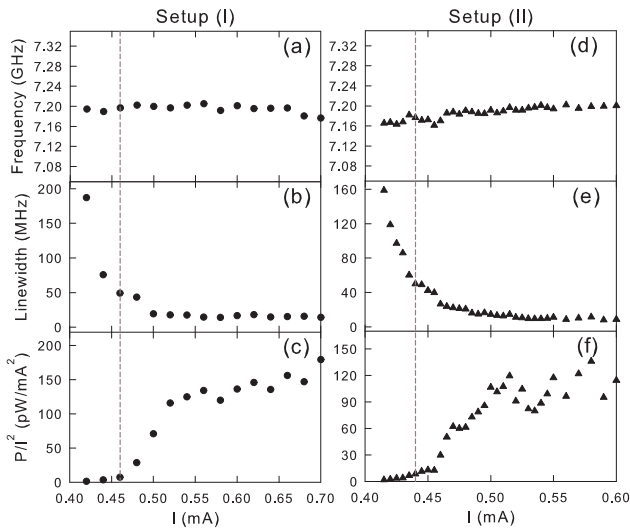}
  \caption{\label{Figure:4}
  Frequency ((a) and (d)), FWHM linewidth $\Delta f$ ((b) and (e)), 
  and normalized power $P\slash I^2$ ((c) and (f)) vs. bias current $I$
  for the setups (I) and (II). 
  In (b), $\Delta f \simeq 14~\mathrm{MHz}$ at $0.58~\mathrm{mA}$. 
  In (e), $\Delta f \simeq 8.6~\mathrm{MHz}$ at $0.56~\mathrm{mA}$. 
  }
\end{figure}

Power spectral density (PSD) of the voltage oscillation signal in the setup (I) 
is shown in Fig.~\ref{Figure:3}. 
The characteristic oscillation mode is observed around $7.2~\mathrm{GHz}$ 
for the bias current $I \ge 0.42~\mathrm{mA}$. 
This mode grows steeply as the bias current increases and 
exhibits narrow linewidth in the high-current region. 
The PSD in the setup (II) is similar to that in the setup (I). 
The dependence of the frequency, 
the full width at half maximum (FWHM) linewidth, 
and the normalized power of the mode on the bias current 
for the setups (I) and (II) is shown in Fig.~\ref{Figure:4}. 
The data are obtained by Lorentzian fits to the spectral peaks 
of the characteristic oscillation modes. 
The broken lines in Fig.~\ref{Figure:4} denote the threshold currents $I_\mathrm{th}$. 
The normalized power below threshold depends on the bias current 
in the way that $(P\slash I^2)^{-1} \propto (I_\mathrm{th}-I)$.\cite{Tiberkevich} 
By using the expression, 
we find that $I_\mathrm{th} \simeq 0.46~\mathrm{mA}$ for the setup (I) and 
$I_\mathrm{th} \simeq 0.44~\mathrm{mA}$ for the setup (II). 
In Figs.~\ref{Figure:4} (a) and (d), 
the frequency shift with the bias current is much smaller than 
that observed in previous experiments for MgO-based STOs\cite{Houssameddine08,Georges}, 
and we cannot judge whether the shift is red or blue. 
In Figs.~\ref{Figure:4} (c) and (f), 
the normalized power grows steeply up to the current 
$I \simeq 0.52~\mathrm{mA}$ from the threshold current. 
For $I > 0.52~\mathrm{mA}$, the normalized power is being saturated.

\begin{figure}[t]
  \includegraphics{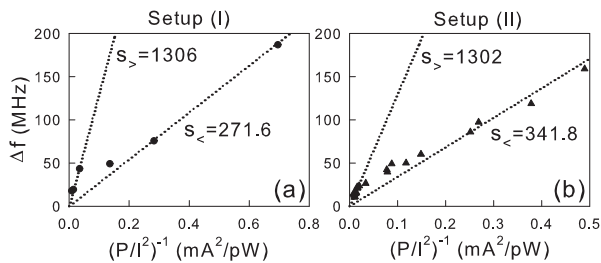}
  \caption{\label{Figure:5}(a) and (b) $\Delta f$ vs. $(P\slash I^2)^{-1}$-plots 
  for the setups (I) and (II). 
  }
\end{figure}

From the data for linewidth $\Delta f$ and normalized power $P\slash I^2$ 
shown in Figs.~\ref{Figure:4}, 
we have obtained the $\Delta f$ vs. $(P\slash I^2)^{-1}$-plots 
shown in Figs.~\ref{Figure:5}. 
We have used the data for the current $I \le 0.52~\mathrm{mA}$. 
As predicted by the theories, 
we observe that linewidth against inverse normalized power plots 
show linear behavior below and above the threshold. 
The appearance of the linear behavior above the threshold 
supports the consideration that 
the variation of $\nu$ with bias current is small 
in the range near threshold. 
By estimating the values of linear slopes $s_{<,>}$ and 
using Eq.~(\ref{eq:relation}), 
we find the coefficient $|\nu| \simeq 2.9$ for the setup (I) 
and $|\nu| \simeq 2.6$ for the setup (II). 
Considering the range of values 
expected by the calculation as shown in Fig.~\ref{Figure:1}, 
these values of the coefficient ($|\nu|\sim 3$) are 
comparatively small. 
The smallness is consistent with 
the flatness of frequency (Figs.~\ref{Figure:4} (a) and (d)) and 
the exhibition of narrow linewidths of $\sim 10~\mathrm{MHz}$ 
in the high-current region (Figs.~\ref{Figure:4} (b) and (e)). 
According to Fig.~\ref{Figure:1}, 
the coefficient $|\nu|$ is expected to have small values 
when an in-plane field is applied to the hard axis, i.e., $\phi_H\sim 80$-$90^\circ$. 
Our results measured in the angle of $\phi_H\sim 40^\circ$ are inconsistent 
with the expectation. 
We consider that the mode with $7.2~\mathrm{GHz}$ peak may be 
a non-uniform mode and so the effective directions and structures of 
in-plane anisotropy for the magnetization composing the mode 
are complicated. 
Micromagnetic study to clarify the oscillation mode is now in progress.

In summary, 
we estimated the coefficient $\nu$ in MgO-based STOs which is a measure of the 
transformation of amplitude fluctuations into phase fluctuations of an oscillator. 
For the device exhibiting narrow linewidth ($\Delta f \sim 10~\mathrm{MHz}$) 
in a high-current region, small value of the coefficient ($|\nu| \sim 3$) was obtained.


\end{document}